\documentclass[12pt,preprint]{aastex}
\usepackage{amsmath}
\begin{document}

\def \evt {OGLE-2003-BLG-095}
\def \bpie {\boldsymbol{\pi}_{\mathrm E}}
\def \pie {\pi_{\mathrm E}}
\def \piee {\pi_{{\mathrm E},e}}
\def \pien {\pi_{{\mathrm E},n}}
\def \rte {\tilde{r}_{\mathrm E}}

\title{The repeating microlensing event OGLE-2003-BLG-095: A plausible case 
of microlensing of a binary source}
\shorttitle{OGLE-2003-BLG-095: Microlensing of a binary source}

\author{Matthew J. Collinge}
\affil{Princeton University Observatory, Princeton, New Jersey 08544\\
collinge@astro.princeton.edu}
\shortauthors{M. J. Collinge}

\begin{abstract}

The apparently repeating microlensing event OGLE-2003-BLG-095 is 
analyzed. Data were obtained from the OGLE Internet 
archive and exist in the public domain. The source is relatively bright, 
with an unmagnified (but possibly blended) 
$I$-band magnitude of 15.58, and the signal-to-noise 
ratio of the data is excellent.
The light curve shows two distinct, smooth peaks
characteristic of a double microlensing event. 
It can be modeled as either (1) microlensing by a binary lens or 
(2) microlensing of a binary source, with the latter model providing 
a statistically superior fit. However due to apparent low-amplitude 
variability of the source, the interpretation is somewhat ambiguous. 
OGLE-2003-BLG-095 is only the second possible case in the literature for 
microlensing of a well-resolved binary source.

\end{abstract}



\section{Introduction}

Gravitational microlensing surveys were originally suggested 
by \citet{pacz86} as a means 
of detecting dark matter in the Galaxy in the form of massive compact 
objects, commonly abbreviated as MACHOs following \citet{grie91}. 
Various efforts were undertaken to search for such objects, including 
the Optical Gravitational Lensing Experiment (OGLE; \citealt{udal92}), 
the MACHO project (e.g., \citealt{alco93}), and several others. 
Currently ongoing 
projects, most notably OGLE and MOA \citep{bond01}, are detecting 
hundreds and dozens of new microlensing events each year, respectively. 
The new events found by these groups are made public in real time 
to maximize scientific gain by enabling follow-up of interesting 
objects by the astronomical community.

In addition to the possibility of detecting dark matter MACHOs, 
microlensing surveys also offer the interesting opportunity 
to study the populations of stellar lenses (which constitute 
at least a significant fraction of all events) and sources. 
Since most stars are located in multiple systems, many microlensing 
events show deviations from the standard Paczy{\'n}ski (point lens, 
point source) 
light curve. For example, some events show the effects of caustic 
crossings \citep{mao91,goul92}; 
this provides clear evidence of lens binarity and 
can be used to constrain the combination of the distances 
to the source and lens, the source-lens 
relative motion, and the physical extent of the source, as 
well as revealing the mass ratio of the lens system.
A small additional fraction of microlensing 
events should show clear signatures of source binarity in multiple 
peaks and color shifts; this fraction was estimated to be 
2--5\% by \citet{grie92}. To date no convincing example of such an event 
has been reported in the literature; the best case is MACHO-96-BLG-4, 
but the light curve of this event can be equally well fit by a binary 
source or binary lens model, with the binary lens 
providing the most natural interpretation \citep{alco00}. 
Several authors 
\citep{domi98,han98,dist00} have 
proposed explanations for the lack of such clear binary source microlensing 
events. Essentially, a variety of different effects including 
blending, unequal luminosity of binary components, and simple coincidence 
conspire to make most binary source microlensing light curves resemble 
single source, single lens events. 
A handful of other events, such as MACHO-96-LMC-2 \cite{alco01}, 
show signatures of 
binary orbital motion of the source; in this particular case, it is not clear 
whether both components of the binary contribute significantly to the 
total source flux. In any case the light curves of such events do not match the 
expectation of having multiple distinct, well-resolved peaks as 
described above. 

This work analyzes the light curve of \evt, which shows two distinct 
microlensing peaks. The light curve can be fit by a binary lens or a 
binary source model; however, the binary source model is 
statistically preferred. 
The light curve is presented in \S\ref{sec:data}, 
\S\ref{sec:model} contains a description of the models and fitting 
procedure, and the results are analyzed in \S\ref{sec:disc}.


\section{OGLE data}
\label{sec:data}

\begin{figure}[htb]
\epsscale{1.0}
\plotone{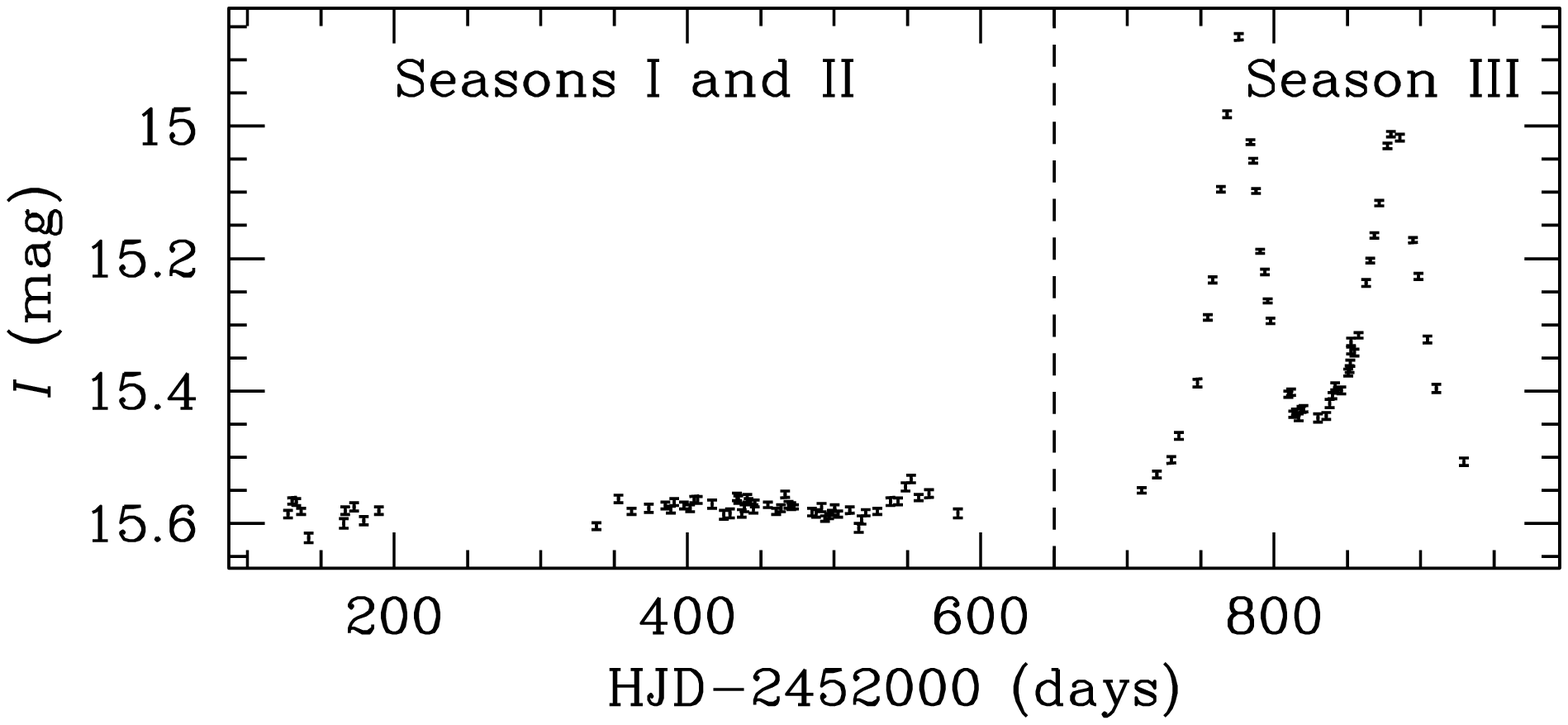}
\figcaption{\label{fig:lc}
$I$-band light curve of \evt\ with nominal uncertainties from the 
OGLE data reduction pipeline. The observing seasons referred to in the 
text are labeled. The source is relatively constant 
during the first two seasons, with the deviations due to microlensing confined 
to the third season. 
}
\end{figure}

An $I$-band light curve of \evt\ was obtained from the OGLE Internet 
archive\footnote{http://sirius.astrouw.edu.pl/\~{}ogle/} (cf.~\citealt{udal03}), 
and is shown in Figure~\ref{fig:lc}. The source is located at 
$(\alpha_{\mathrm J2000},\delta_{\mathrm J2000})=(17$:59:03.04, $-27$:11:$19.8)$; 
a finding chart is also available from the OGLE web site.
The data set consists of 109 observations from three observing seasons 
spanning a period of 2.2~yr, from August 2001 through 17 October 2003.
The light curve from the 2001 and 2002 observing seasons 
(seasons I and II hereafter) 
is relatively constant (to within $\pm 3$\% excluding the highest point 
and the lowest point), with a mean $I$-band magnitude of 15.58. 
However, due to the high stellar surface number density in the field, 
some blending seems likely.
The light curve from the 
2003 season (III) shows two very distinct smooth peaks, with maximum 
magnification factors of $\approx 1.9$ and $\approx 1.7$. For both peaks, the 
rise and fall in source intensity are well covered by observations.

In order to simplify the fitting procedure, the light curve 
was converted to units of relative flux by adopting the mean 
intensity from seasons I and II
as the base flux level. Thus the mean relative flux from seasons I and II is unity 
by construction. Even for relatively bright sources like the one in consideration, 
the photometric error estimates reported by the OGLE data reduction pipeline 
(cf.~\citealt{udal02}) are typically somewhat too small to account for the 
observed scatter in the light curves of sources that do not vary intrinsically 
(B.~Paczy{\'n}ski 2004, private communication). In the \evt\ light 
curve from seasons I and II, the standard deviation $s$ in flux is larger 
than the mean quoted error $\bar{\sigma}$ by a factor $s/\bar{\sigma}=2.65$, 
a ratio which nevertheless is significantly 
too large to be readily explained by underestimation of uncertainties. 
This is probably an indication that the source (or at least one component of 
it, if it is a blend) is variable at the level of at least a few percent.
A careful inspection of the season II light curve 
indicates that variations are correlated from night to night.
However, no periodic signal 
is present at a level of significance above 50\% in a Lomb-Scargle 
normalized periodogram (cf.~\citealt{pres92,schw98}) 
computed for seasons I and II. 

Such non-periodic variability substantially complicates the modeling 
process, causing greatest confusion for model effects that 
produce variations with similar amplitudes 
and time scales. See \S\ref{sec:disc} for 
further discussion. To aid in interpreting fit results, the 
errors were scaled so that $\chi^2$ per degree of freedom $\nu$ (DOF), 
also known as the reduced $\chi^2$ and denoted $\chi^2_{\nu}$, is unity for a 
constant fit to the light curve from seasons I and II. This required 
scaling all relative flux errors by a factor of 2.49. The scaling factor 
is similar to the ratio $s/\bar{\sigma}$ quoted in the previous paragraph; the 
fact that these two quantities differ could either be attributed to a non-Gaussian 
error distribution (which seems likely given the apparent systematic variations) or 
to the limited number of data points.


\section{Modeling the light curve}
\label{sec:model}

Since the intensity of the source during seasons I and II was taken to 
be constant, microlensing models were applied to the 50 data points 
from season III only. Two classes of models were considered: 
(1) models with a coplanar binary lens consisting of two point masses, 
in which the source was taken to be a small (but finite) disk of 
uniform surface brightness (``binary lens'' or ``double lens'' models); 
(2) models with a single point mass lens and two distinct point 
sources (``binary source'' or ``double source'' models).
The source was treated as finite in the former case due to the 
possibility of caustic crossings; although there is no direct 
evidence of such in the \evt\ light curve, the possibility was 
considered in order not to restrict the exploration of parameter space 
(e.g., a case in which a brief caustic crossing event occurs in a 
gap between observations). In the latter case, such caution was not 
needed since the smooth, low-magnification light curve precludes 
any possibility of very small (close to zero) impact parameters.

The light curve of \evt\ suggests that the two binary components (whether of the 
lens or of the source) are separated by an angular distance of at 
least twice the angular Einstein radius $\theta_{\mathrm E}$. For 
a typical value of $\theta_{\mathrm E}\approx 0.5$~milli-arcseconds (mas) 
and a distance to the binary (lens or source) of at least a few kiloparsecs (kpc), 
which is very likely for an event in the direction of the Galactic bulge, 
the implied projected separation of binary components is 
several Astronomical Units (AU). Assuming the masses to be not too different 
from a Solar mass ($M_{\odot}$), the minimum orbital period of the system 
is several years.
Therefore, both classes of models were restricted to the static case; that is, 
any relative motion of binary components (lens or source) was neglected. 

\subsection{Parallax effect}
\label{sec:parallax}

For both classes of models, two parameters are needed to characterize 
the vector microlens parallax $\bpie$ (e.g., \citealt{goul00}). One 
parameter can be chosen to be the parallax magnitude $\pie$, whose inverse 
gives the projected size $\rte$ of 
the Einstein radius in the observer plane. A second parameter $\psi$ then 
gives the lens-source proper motion angle on the sky (measured 
counter-clockwise from North). 
Following the method recently proposed by \citet{goul03}, all parallax fitting 
is done in geocentric reference frames defined with respect to the position and 
velocity of Earth at times very close to one of the light curve peaks. 

\subsection{Binary lens model}
\label{sec:modellens}

Binary lens models are characterized by eight parameters, in addition 
to the two components of the microlens parallax.
Four parameters describe the trajectory of the lens relative to the source: 
the time $t_0$ of closest approach to the more massive binary 
component (corresponding roughly to one of the two event maxima), 
the Einstein time scale $t_{\mathrm E}$, the angle $\phi$ between the lens line 
of centers and the lens trajectory with respect to the source position, 
and the impact parameter $u_0$ to the more massive binary component.
The remaining four parameters describe the lens and source properties: 
the binary mass ratio $q$ and separation $a$, the 
fraction $f_{\mathrm{bl}}$ of the baseline intensity due to the source, 
and the ratio of the angular source radius to the angular Einstein radius, 
$\xi_{\star,\mathrm E}\equiv \theta_{\star}/\theta_{\mathrm E}$.
All length scales ($a, u_0$) are in terms of the Einstein 
radius $r_{\mathrm E}$ of a lens with mass equal to the total mass of the binary.

For events in which it has been possible to measure or estimate 
the ratio $\xi_{\star,\mathrm E}$ due to caustic crossings, 
typical values for microlensing 
toward the Galactic bulge are found to be on the order of 0.001--0.01 
\citep{alco00, jaro02}. 
In the absence of caustic crossings and at low magnification, as in 
the case of \evt, binary lens model light curves are very insensitive to the 
precise value of $\xi_{\star,\mathrm E}$ provided it is smaller than a 
few tens of percent. Hence this parameter is poorly constrained by 
the available data, and its value is fixed at 0.01 in the subsequent analysis.

For a binary lens the model flux can conveniently be found numerically 
for a given lens-source configuration by finding the 
(multiple) image locations for a set of points on the source boundary
(e.g., \citealt{goul97}).
This procedure gives 3--5 new sets of points defining the image boundaries, 
depending on whether the source is outside, crossing, or inside a caustic.
Since the surface brightness of the source is taken to be constant, the 
magnification is given by the ratio of the sum of the (appropriately 
signed) areas enclosed by the image boundaries to the area of the source.

\subsection{Binary source model}
\label{sec:modelsource}

The static binary source model is constructed from a 
linear combination of two point lens, point source models 
with Einstein time scales and vector parallaxes 
constrained to be the same 
for each source component. 
This model has seven parameters in addition to parallax: 
two for the fraction of the baseline intensity due to each source 
($f_{{\mathrm{bl}},i}$, where $i=1,2$), three for the trajectory 
($u_0,t_0,\phi$ as in the previous case except that the reference 
point is defined by the closest approach of the lens to the source 
responsible for the first peak in the light curve), source 
separation $a$, 
and the Einstein time scale $t_{\mathrm E}$.

The model flux $m(t)$ at time $t$ can be expressed as 
\begin{equation}
\label{eq:modelflux}
m(t) = 1+\sum_{i=1}^{2} f_{{\mathrm{bl}},i}[A(u_i)-1]
\end{equation}
where
\begin{equation}
\label{eq:mag}
A(u)=\frac{{u^2+2}}{u(u^2+4)^{1/2}}
\end{equation}
gives the total magnification for a point source, point lens system with 
separation $u$ (in units of $r_{\mathrm E}$). In this parametrization, the 
condition $f_{{\mathrm{bl}},1}+f_{{\mathrm{bl}},2}\leq 1$ 
must clearly be met in order for 
the model to be physically meaningful.

\subsection{Fitting procedure}
\label{sec:modelfitproc}

The $\chi^2$--minimization was performed using MINUIT, 
a function minimization package available as part of the 
CERN program library\footnote{http://cernlib.web.cern.ch/cernlib/index.html}.
For each class of models, the fitting algorithm was initialized in 
several different realizations with different sets of initial parameter 
values. All such sets were chosen to produce an initial model light curve 
that was qualitatively similar to the observed one 
(two well-separated smooth peaks with approximately the correct amplitude 
and width). For the 
binary lens model, four different initial source trajectories were chosen: 
two in which the source encountered the more massive lens first; two in 
which the source encountered the less massive lens first. For each of these 
pairs, one trajectory was chosen to pass in between the two lenses, and 
one was chosen not to pass in between the two lenses. The initial values 
of the other parameters were chosen appropriately to reproduce qualitatively 
the light curve shape for each of these cases. For the binary 
source model, an equivalent variety of initial parameter guesses were used, 
subject again to the requirement that the model light curve be qualitatively 
similar to the observations.
After the best fit solutions were obtained for models in which the 
parallax was fixed at zero, 
full nine parameter fits (including parallax) 
were performed using the previous solutions as the initial guesses.

\subsection{Fitting results}
\label{sec:modelfitres}

\begin{figure}[htb]
\epsscale{1.0}
\plotone{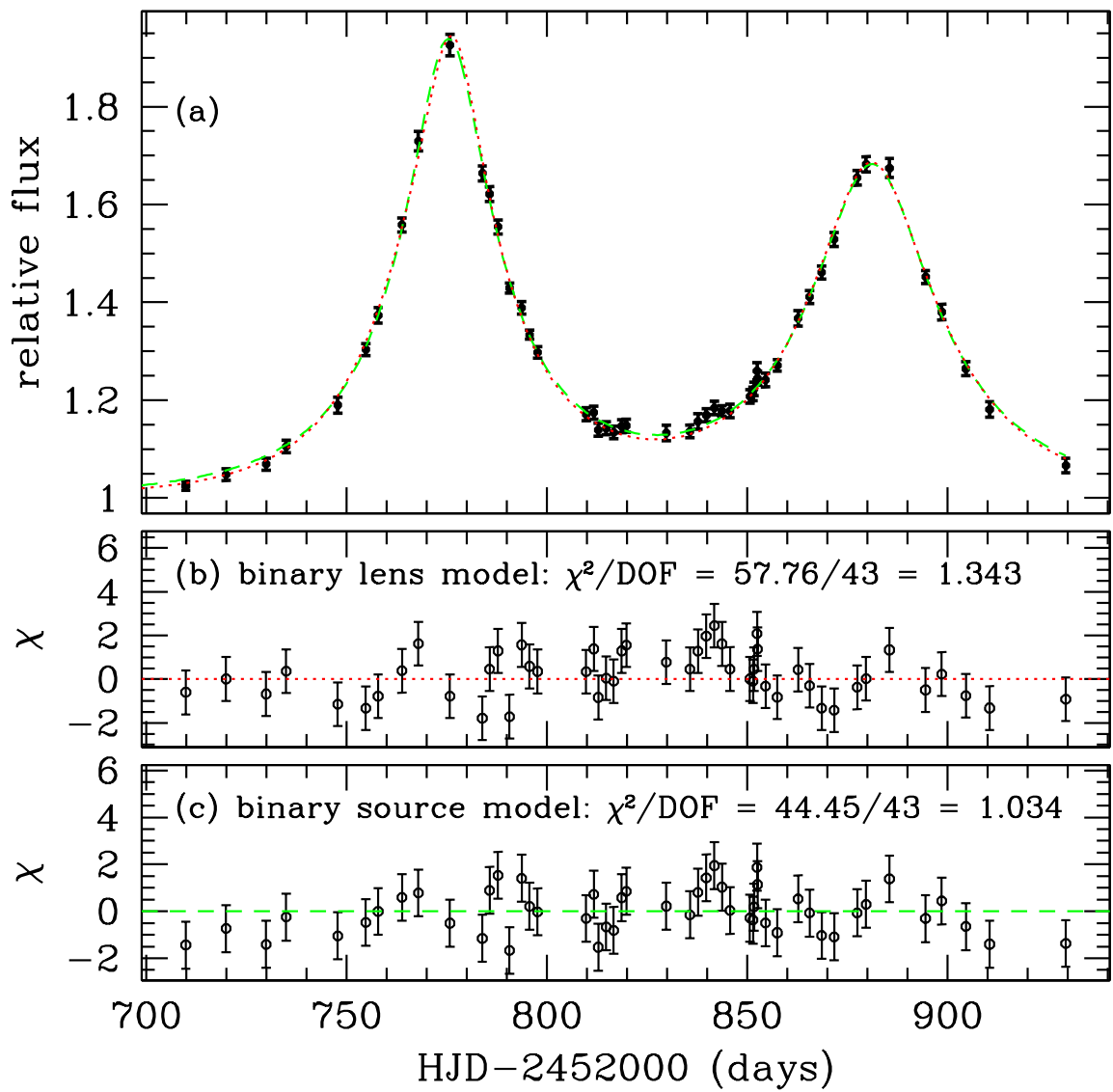}
\figcaption{\label{fig:fit}
Panel (a): Best fit no-parallax binary lens (dotted curve) 
and binary source (dashed curve) models for the light curve of \evt. 
To the eye, there is little difference 
in the qualities of the two fits. Panel (b): Fit residuals in units of 
sigmas for the best fit binary lens model (not including parallax). 
Panel (c): Same as panel (b), but for the 
binary source model. The residuals in both of the lower panels display 
a somewhat systematic character.
}
\end{figure}

The best fit no-parallax model light curves and fit residuals 
are shown in Figure~\ref{fig:fit}. 
Tables~\ref{tab:doublelens} and~\ref{tab:doublesource} contain the 
best fit results for the binary lens and binary source models, respectively. 
For the same number of DOF, the 
no-parallax double source model is preferred over the no-parallax double 
lens model by a highly significant 
margin of $\Delta\chi^2=13.3$, and furthermore this model satisfies 
the criterion that $\chi^2_{\nu}\approx 1$ (with the caveat that the 
overall error scaling is rather uncertain). 
However, the lower panels of Figure~\ref{fig:fit}
both show evidence of systematic residuals, indicating that unmodeled 
complexity is present in both fits. This is not at all surprising 
in light of the evidence for 
low-amplitude source variability mentioned in \S\ref{sec:data}, and 
considering that parallax has been neglected.

Including the parallax effect adds two more free model parameters and 
results in $\Delta\chi^2=-25.7$ and 
$\Delta\chi^2=-16.5$ for the binary lens and binary source models, 
respectively. These large $\Delta\chi^2$ values formally indicate that 
the parallax effect is detected with a high level of statistical significance. 
The inclusion of parallax brings both models into the 
$\chi^2_{\nu}<1$ regime, for the adopted error scaling. It also 
reduces the difference in the statistical 
quality of the two model fits, though the double source model remains 
superior by $\Delta\chi^2 = 4.15$ for the same number of DOF. 

It is worthwhile to note that although the overall 
error scaling is rather arbitrary as discussed in 
\S\ref{sec:data} (and hence the precise value of $\chi^2$ 
is far from rigorous), $\Delta \chi^2$ is still a 
meaningful quantity. The low values of $\chi^2_{\nu}=0.783$ and 
$\chi^2_{\nu}=0.682$ respectively obtained 
for the best fit double lens and double source models indicate that the errors 
have probably been scaled by too large a factor, resulting in values of 
$\chi^2$ that are systematically low. In principle, the errors could be 
rescaled so that $\chi^2_{\nu}=1$ for the best fitting model, but this 
does not seem warranted, especially in view of the complications 
introduced by the likely source variability. Regardless, if the errors had been 
scaled by a smaller factor, $\Delta \chi^2$ values would increase 
correspondingly. 

Additionally, there are a number of degeneracies in the problem that are 
important to consider. In the absence of parallax, that is, when the apparent 
lens-source relative trajectory is (modeled as being) strictly linear, 
there are two fully degenerate 
binary lens models and four fully degenerate binary source 
models. The degenerate models differ only by the sign of the impact 
parameter $u_0$ and the orientation angle $\phi$. There 
are only two equivalent binary lens solutions because trajectories on 
which the two lens components pass on opposite sides of the source 
produce fundamentally different light curves (due to proximity to caustics) 
from trajectories on which the two lens components 
pass on the same side of the source. The same does not hold for the binary 
source models, and hence in this case there exist four degenerate trajectories. 

These degeneracies are somewhat alleviated when the parallax effect is included. 
For the binary lens model the degeneracy is broken; the two solutions 
separate by $\Delta\chi^2=2.79$ (only the preferred solution is given in 
Table~\ref{tab:doublelens}). For the binary source model the degeneracy 
is largely broken as well; only the solution with the lowest $\chi^2$ is 
given in Table~\ref{tab:doublesource}. The three solutions not included 
in Table~\ref{tab:doublesource} all fall within the range $1<\Delta\chi^2<2$ 
of the preferred solution. Regardless, the differences among these 
near-degenerate solutions 
are largely immaterial, since the best fit parameters (neglecting the sign 
of the impact parameter $u_{0}$ and the orientation angle $\phi$) lie within 
or very near the 1-$\sigma$ error ranges quoted in 
Tables~\ref{tab:doublelens} and~\ref{tab:doublesource}. 
For all the solutions, the parallax magnitudes $\pie$ are consistent within 
1-$\sigma$ with the values from the best fit solutions.

\begin{figure}[htb]
\epsscale{1.0}
\plotone{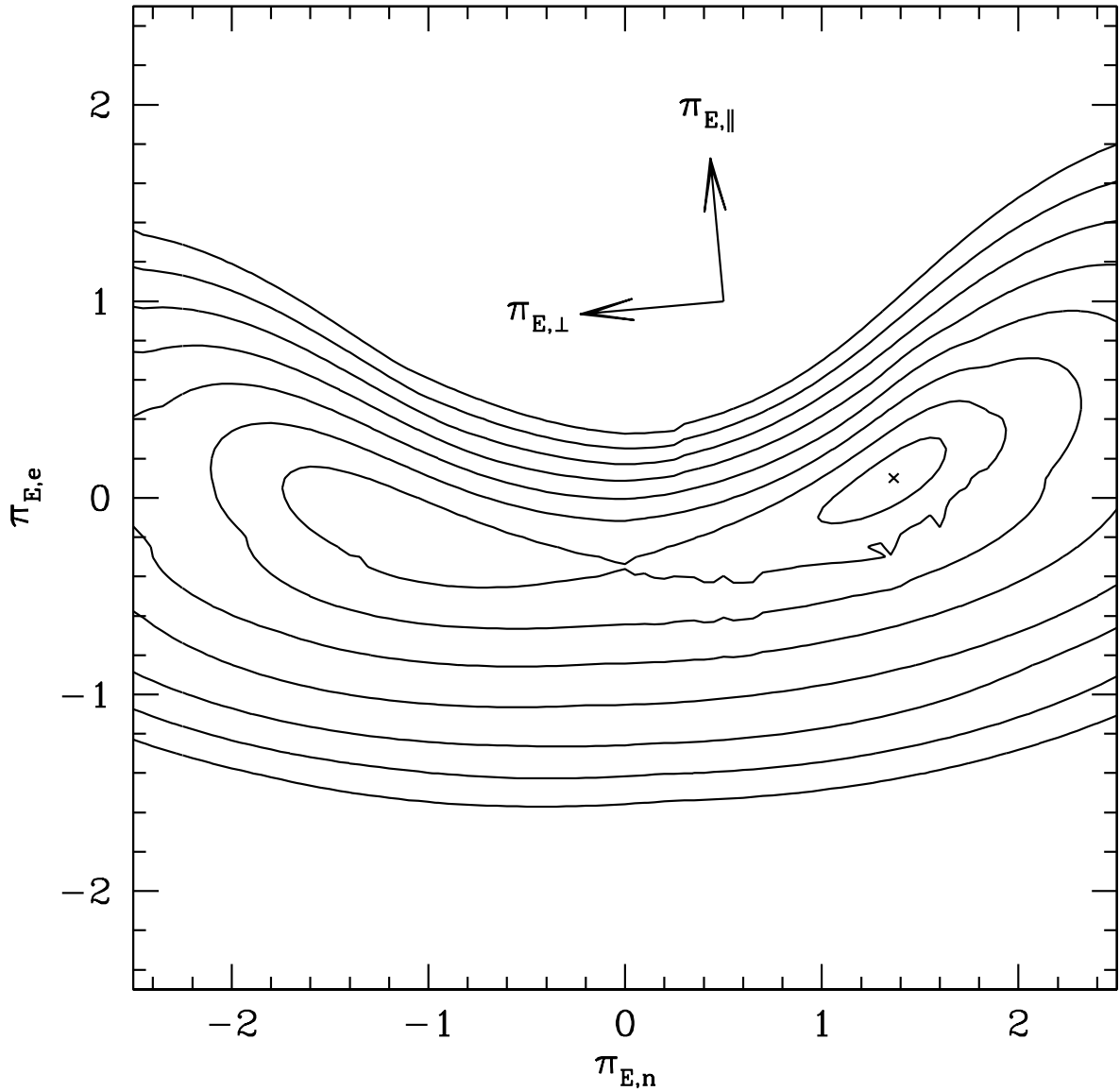}
\figcaption{\label{fig:fitcon}
$\chi^2$ contours in the $\bpie$ plane for the double source model. 
`X' marks the best fit solution; contours corresponding to 
$\Delta\chi^2=1,4,9,16,25,36,49,64$ are shown. 
The subscripts $n,e$ refer to the North and East 
components of the parallax vector. 
The symbols $\parallel$ and $\perp$ refer to the projected direction of the 
Earth's acceleration vector at the time of the first event maximum (which is 
the reference point of the geocentric frame in which parallax fitting is done).
Although the contours are elongated in the expected direction, 
no fully degenerate solution 
is found as in \citet{goul03}.
}
\end{figure}

There is a further possible degeneracy discovered recently by \citet{goul03}. 
This ``jerk-parallax'' degeneracy can produce multiple parallax solutions 
for a given sign of the impact parameter. In an effort to find any such 
solutions, a detailed search of 
the $\bpie$ plane was performed for the 
double source model (which is more computationally tractable than 
the double lens model). The results are shown in Figure~\ref{fig:fitcon}. 
No fully degenerate solution is present,
although the contours are highly elongated in a direction 
perpendicular to the projected acceleration vector of Earth at the time 
of the first event maximum, when the geocentric reference frame is defined 
(as expected for this degeneracy).
Numerical searches for other parameter combinations that produce 
similar trajectories were also performed (by solving eqs.~12--14 
from \citealt{goul03}) for both classes of models. No further degenerate 
solutions were discovered.


\section{Discussion}
\label{sec:disc}

\begin{figure}[htb]
\epsscale{1.0}
\plotone{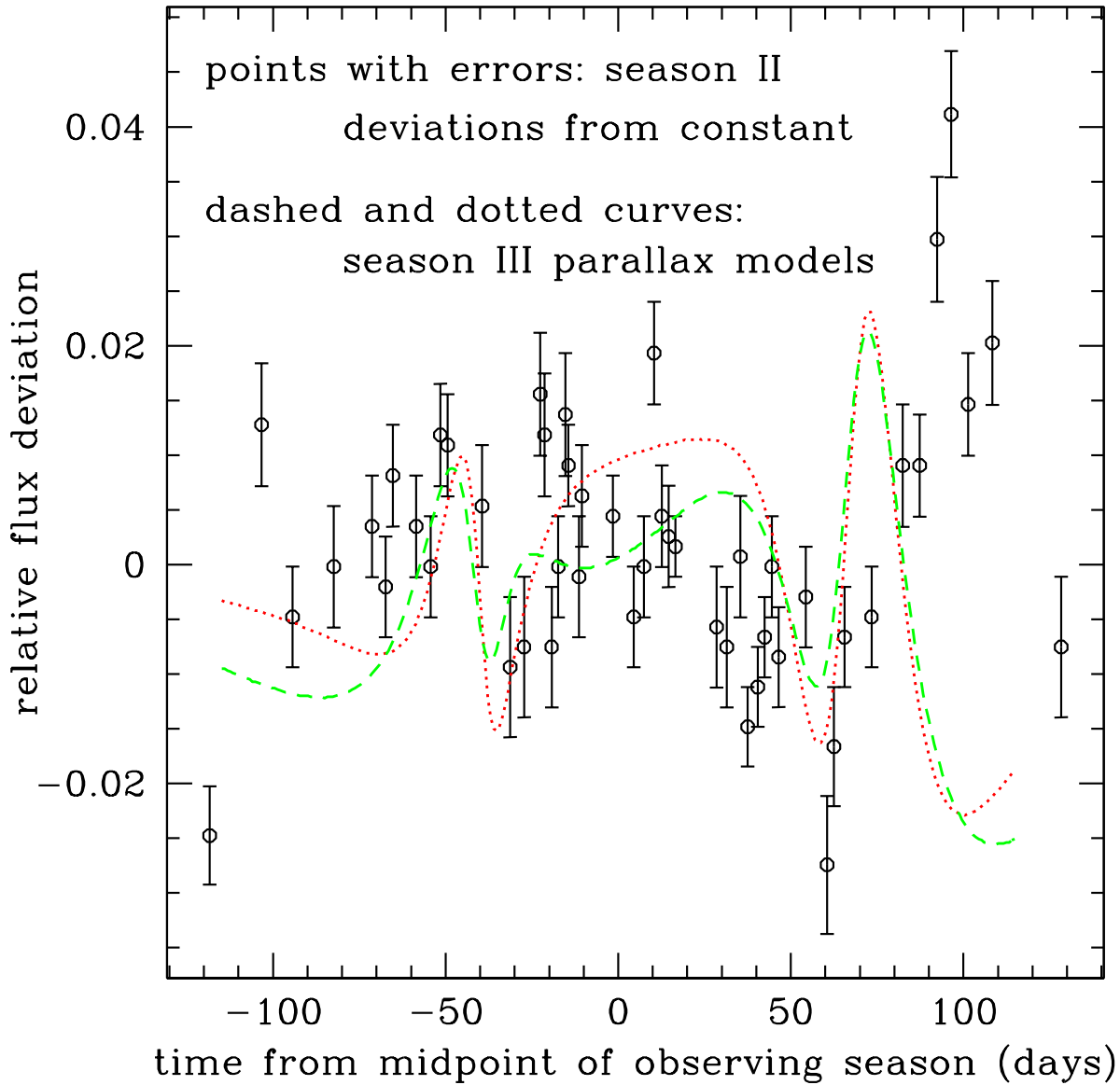}
\figcaption{\label{fig:modeldiff}
The curves do not represent fits to the data.
Points with errorbars show deviations from constant relative flux from season II 
with nominal OGLE data reduction pipeline uncertainties. 
The dotted line is the binary lens parallax-included fit minus no-parallax fit from 
season III; the dashed line is the same, but for the binary source model. 
The qualitative similarities between the season II light curve and the season III 
``model-difference'' curves are striking, especially considering that they refer to 
different observing seasons. 
The time scales for variation are similar, 
and notably, the amplitude of deviations from constant 
flux during season II is comparable to or 
greater than the differences between the 
parallax-included and no-parallax fits, even though no scaling has 
been applied to account for the fact that the source was brighter during 
season III (and thus relative flux deviations due to intrinsic source 
variability should be magnified as well).
}
\end{figure}

An extremely relevant question is whether the parallax effect can 
be disentangled from the likely variability of the source (regardless 
of whether the variability is inherent or due to 
a hitherto unrecognized systematic effect). To 
address this issue, Figure~\ref{fig:modeldiff} compares 
the deviations from constant flux in the season II light curve 
with the magnitude of the effect of parallax on the model light curves 
for season III. It is apparent that the modulations 
of the model light curves due to the parallax effect have 
similar amplitudes and variation time scales to the observed 
variability of the source, rendering the parallax measurements 
rather untrustworthy.

Further support for the notion of ``contaminated'' parallax measurements 
comes from examining the implications of the best fit parallax-included 
models. There is a simple relationship between the projected Einstein 
radius $\rte$, the lens mass $M$, and the lens-source relative distance 
$D_{\mathrm{rel}}\equiv (D_l^{-1}-D_s^{-1})^{-1}$, where $D_l$ and $D_s$ 
are the distances from the observer to the lens and source; this 
relation is given by
\begin{equation}
\label{eq:drel}
D_{\mathrm{rel}}=\frac{c^2 \tilde{r}^2_{\mathrm E}}{4GM}
\end{equation}
\citep{goul00}. It is plausible to assume the source to be located in 
the Galactic bulge at a distance of approximately 8~kpc 
and to assume the lens to have a total mass of approximately one 
Solar mass. Substituting the measured values of 
$\pie$ from the best fit binary lens and binary source models, one 
obtains $D_{\mathrm{rel}}\approx 0.25$~kpc and 
$D_{\mathrm{rel}}\approx 0.06$~kpc, respectively.
In both cases $D_{\mathrm{rel}}^{-1}\gg D_s^{-1}$, indicating that 
$D_l\approx D_{\mathrm{rel}}$. Even accounting for the substantial 
additional uncertainties introduced by assuming the lens mass and source 
distance, such a nearby lens seems unlikely.

Since the apparent variability of the source limits the degree to 
which the binary lens and binary source models can be statistically 
distinguished, a definitive interpretation of \evt\ will probably require 
further observations. At the least, future OGLE observations of the 
source will allow better constraints on its variability. A simple 
(though not necessarily straightforward) test which might clearly confirm the 
binary source model could be performed through moderate resolution optical 
spectroscopy to see if the source exhibits double absorption lines or radial 
velocity variations. An alternative possibility would be to obtain 
deep, high resolution imaging of the field to see if the lens can be 
observed directly.


\section{Conclusion}
\label{sec:conc}

The light curve of \evt\ shows two well-separated smooth peaks that can 
be modeled as either microlensing by a binary lens or microlensing of 
a binary source. Both models are physically plausible and can possibly be 
tested by future observations of the source and/or lens.
The binary lens model provides a more 
familiar explanation of the event since such phenomena have been observed 
many times in the past, while the binary source 
model is preferred on statistical grounds. Given the apparent 
low-amplitude variability 
of the source (or at least one component of it) however, statistical 
distinctions of the magnitude that separate the two models are somewhat 
suspect, especially considering the improbably large microlens parallaxes 
indicated by the best fit models. Despite these complications, 
\evt\ remains a plausible candidate for microlensing of a 
well-resolved binary source -- one of only two such observed to date.


\acknowledgements
The author wishes to thank B.~Paczy{\'n}ski and T. Sumi for helpful comments and 
suggestions, M. Jaroszy{\'n}ski for providing routines (originally 
developed by S.~Mao) to calculate binary lens light curves, 
A. Gould for helpful discussion, and the OGLE collaboration 
for making microlensing data publicly available in real time. This research 
was supported in part by a NDSEG Fellowship award to the author.




\clearpage
\begin{deluxetable}{cccc}
\tablecolumns{4} 
\tablewidth{0pc} 
\tablecaption{\label{tab:doublelens}Parameters of best fit double lens model}
\setcounter{table}{0}
\tablehead{\colhead{Quantity} & \colhead{No-parallax}    & \colhead{Parallax included} & \colhead{Units} \\
           \colhead{}         & \colhead{best fit value} & \colhead{best fit value}    & \colhead{}}
\renewcommand{\arraystretch}{1.2}
\startdata
 $\pie$                   & 0.0\tablenotemark{a}   & $0.7\pm0.2$              & AU$^{-1}$ \\
 $\psi$                   & 0.0\tablenotemark{a}   & $-0.2^{+0.2}_{-0.4}$     & rad \\
 $q$                      & $1.28\pm0.05$          & $1.3\pm0.1$              & \nodata \\
 $a$                      & $2.90\pm0.07$          & $3.1\pm0.1$              & $r_{\mathrm E}$ \\
 $\phi$                & $0.048^{+0.005}_{-0.004}$ & $0.25^{+0.06}_{-0.07}$   & rad \\
 $u_0$                    & $-0.49\pm 0.02$        & $-0.51^{+0.03}_{-0.02}$  & $r_{\mathrm E}$ \\
 $t_0$                    & $884.3\pm 0.3$         & $884.5\pm0.5$            & days\tablenotemark{b} \\
 $t_{\mathrm E}$          & $39\pm 1$              & $38\pm2$                 & days \\
 $f_{\mathrm{bl}}$         & $0.92^{+0.06}_{-0.05}$ & $0.95^{+0.05}_{-0.08}$  & \nodata \\
 $\xi_{\star,\mathrm E}$  & 0.01\tablenotemark{a}  & 0.01\tablenotemark{a}    & $\theta_{\mathrm E}$ \\
 \nodata                  & \nodata                & \nodata                  & \nodata \\
 $\chi^2$                 & 57.76                  & 32.10                    & \nodata \\
 DOF                      & 43                     & 41                       & \nodata \\
 $\chi^2_{\nu}$           & 1.34                   & 0.783                    & \nodata \\
\enddata
\renewcommand{\arraystretch}{1.0}
\tablecomments{Parameter values for the no-parallax fit refer to a heliocentric frame; those for the parallax-included fit refer to a geocentric frame. Uncertainties are quoted at the $\Delta\chi^2=1$ level.}
\tablenotetext{a}{fixed}
\tablenotetext{b}{HJD-2452000}
\end{deluxetable}
\begin{deluxetable}{cccc}
\tablecolumns{4} 
\tablewidth{0pc} 
\tablecaption{\label{tab:doublesource}Parameters of best fit double source model}
\tablehead{\colhead{Quantity} & \colhead{No-parallax}    & \colhead{Parallax included} & \colhead{Units} \\
           \colhead{}         & \colhead{best fit value} & \colhead{best fit value}    & \colhead{}}
\renewcommand{\arraystretch}{1.2}
\startdata
 $\pie$                   & 0.0\tablenotemark{a}   & $1.4^{+0.3}_{-0.4}$      & AU$^{-1}$ \\
 $\psi$                   & 0.0\tablenotemark{a}   & $0.07^{+0.12}_{-0.20}$   & rad \\
 $f_{{\mathrm{bl}},1}$    & $0.28\pm0.03$          & $0.34\pm0.04$            & \nodata \\
 $u_{0}$                  & $0.23\pm0.02$          & $0.28\pm0.03$            & $r_{\mathrm E}$ \\
 $t_{0}$                  & $775.4\pm0.1$          & $775.3\pm0.2$            & days\tablenotemark{b} \\
 $t_{\mathrm E}$          & $44\pm2$               & $41^{+3}_{-2}$           & days \\
 $f_{{\mathrm{bl}},2}$    & $0.33\pm0.04$          & $0.28^{+0.07}_{-0.05}$   & \nodata \\
 $a$                      & $2.4\pm0.1$            & $2.3^{+0.2}_{-0.1}$      & $r_{\mathrm E}$ \\
 $\phi$                   & $-0.046\pm0.005$       & $0.03^{+0.03}_{-0.04}$   & rad \\
 \nodata                  & \nodata                & \nodata                  & \nodata \\
 $\chi^2$                 & 44.45                  & 27.95                    & \nodata \\
 DOF                      & 43                     & 41                       & \nodata \\
 $\chi^2_{\nu}$           & 1.03                   & 0.682                    & \nodata \\
\enddata
\renewcommand{\arraystretch}{1.0}
\tablecomments{Parameter values for the no-parallax fit refer to a heliocentric frame; those for the parallax-included fit refer to a geocentric frame defined by the position and velocity of Earth near the time $t_0$. Uncertainties are quoted at the $\Delta\chi^2=1$ level.}
\tablenotetext{a}{fixed}
\tablenotetext{b}{HJD-2452000}
\end{deluxetable}

\end{document}